\documentstyle[12pt,epsfig,aaspp4]{article}

\received{ }
\accepted{ }

\lefthead{Lessard et al.}
\righthead{Search for Pulsed TeV Gamma-ray Emission from the Crab Pulsar}
\newcommand{\degr}{^\circ}

\begin{document}

\title{Search for Pulsed TeV Gamma-ray Emission from the Crab Pulsar}

\author{
R.W.Lessard\altaffilmark{1}, 
I.H.Bond\altaffilmark{2}, 
S.M.Bradbury\altaffilmark{2}, 
J.H.Buckley\altaffilmark{3}, 
A.M.Burdett\altaffilmark{2,4}, 
D.A.Carter-Lewis\altaffilmark{5}, 
M.Catanese\altaffilmark{5},
M.F.Cawley\altaffilmark{6},
M.D'Vali\altaffilmark{2},
D.J.Fegan\altaffilmark{7}, 
J.P.Finley\altaffilmark{1}, 
J.A.Gaidos\altaffilmark{1}, 
G.H.Gillanders\altaffilmark{8},
T.Hall\altaffilmark{1}, 
A.M.Hillas\altaffilmark{2},
F.Krennrich\altaffilmark{5}, 
M.J.Lang\altaffilmark{8},
C.Masterson\altaffilmark{7}, 
P.Moriarty\altaffilmark{9},
J.Quinn\altaffilmark{7}, 
H.J.Rose\altaffilmark{2},
F.W.Samuelson\altaffilmark{5},
G.H.Sembroski\altaffilmark{1}, 
R.Srinivasan\altaffilmark{1}, 
V.V.Vassiliev\altaffilmark{4},
T.C.Weekes\altaffilmark{4}}

\altaffiltext{1}{Department of Physics, Purdue University, West Lafayette, 
IN 47907}
\altaffiltext{2}{Department of Physics, University of Leeds, Leeds, LS2 9JT,
U.K.}
\altaffiltext{3}{Department of Physics, Washington University, St. Louis,
MO 63130}
\altaffiltext{4}{Whipple Observatory, Harvard-Smithsonian CfA, P.O. Box 97, 
Amado, AZ 85645-0097}
\altaffiltext{5}{Department of Physics and Astronomy, Iowa State University, 
Ames, IA 50011}
\altaffiltext{6}{Department of Physics, National University of Ireland, 
Maynooth, Co. Kildare, Ireland}
\altaffiltext{7}{Department of Exp. Physics, University College Dublin,
Belfield, Dublin 4, Ireland}
\altaffiltext{8}{Department of Physics, National University of Ireland, 
Galway, Ireland}
\altaffiltext{9}{School of Science, Galway-Mayo Institute of Technology, 
Galway, Ireland}
\authoremail{lessard@physics.purdue.edu}
\authoraddr{Rodney W. Lessard, Department of Physics, Purdue University, West
Lafayette, IN 47907}

\begin{abstract} 
We present the results of a search for pulsed TeV emission from the
Crab pulsar using the Whipple Observatory's 10~m gamma-ray telescope.
The direction of the Crab pulsar was observed for a total of
73.4~hours between 1994 November and 1997 March.  During this period
the Whipple 10~m telescope was operated at its lowest energy threshold
to date.  Spectral analysis techniques were applied to search for the
presence of a gamma-ray signal from the Crab pulsar over the energy
band 250~GeV to 4~TeV. We do not see any evidence of the 33~ms
pulsations present in other energy bands from the Crab pulsar.  The
99.9\% confidence level upper limit for pulsed emission above 250~GeV
is derived to be $4.8\times10^{-12}{\rm cm}^{-2}{\rm s}^{-1}$ or
$<$3\% of the steady flux from the Crab Nebula. These results imply a
sharp cut-off of the power-law spectrum seen by the EGRET instrument
on the {\em Compton Gamma-Ray Observatory}.  If the cut-off is
exponential, it must begin at 60~GeV or lower to accommodate these
upper limits.
\end{abstract}


\setcounter{footnote}{0}
\section{Introduction}
\label{section:intro}
The Crab pulsar/Nebula system is one of the most intensely studied
astrophysical sources with measurements throughout the electromagnetic
spectrum from the radio to the TeV energy band. In most regions of the
spectrum, the characteristic 33~ms pulsations of the pulsar are
clearly visible. The pulse profile is unique amongst known pulsars in
that it is aligned from radio to gamma-ray energies. The study of the
pulsed emission in different energy ranges is of considerable
importance to understanding the underlying emission mechanisms (e.g.,
\cite{eikenberry97}). The EGRET instrument on the {\em Compton
Gamma-Ray Observatory (CGRO)} has shown that there is pulsed gamma-ray
emission from the pulsar up to at least 10~GeV
(\cite{ramanamurthy95}). Current imaging atmospheric Cherenkov
telescopes have firmly established the Crab Nebula as a steady source
of gamma rays from 300~GeV to 50~TeV (\cite{hillas98};
\cite{tanimori98}). However, these observations have not detected any
significant modulation of this TeV signal at the period of the
pulsar. In contrast to these reports, other groups have reported TeV
emission modulated at the 33 ms period of the Crab pulsar. Some of
these reports have been associated with episodic activity
(\cite{gibson82}; \cite{bhat86}; \cite{acharya92}). A persistent
pulsed signal from the Crab pulsar was reported by the Durham group
(\cite{dowthwaite84}). However this has not been confirmed by more
sensitive observations which show that less than 5\% of the total very
high energy (VHE) flux is pulsed (\cite{weekes89}; \cite{reynolds93};
\cite{goret93}). At ultra-high energies, the CASA-MIA experiment does
not find any statistically significant evidence for pulsed gamma-ray
emission at the Crab pulsar period, on an interval of one day or
longer, based on the analysis of data recorded during the interval
1990 March to 1995 October (\cite{borione97}).

Pulsed emission from the Crab pulsar at IR energies and above is
generally believed to originate in the magnetosphere of the system far
from the stellar surface. In each of the two models which address the
pulsed gamma-ray emission in detail, the outer gap model
(\cite{cheng86}; \cite{romani96}) and the polar cap model
(\cite{daugherty82}), the high energy flux arises from curvature
radiation of pairs as they propagate along the open field lines of the
magnetosphere. The specific details of the pulse shapes in different
pulsars are explained by the line of sight geometry of the observer
relative to the spin and magnetic axes of the rotating neutron star in
these models. The energy at which the pulsed flux begins to cut-off
and the detailed spectral shape of the cut-off can help to distinguish
between the two models. Given the detection of pulsations out to
10~GeV by EGRET (\cite{ramanamurthy95}) and the restrictive upper
limits above 300~GeV (\cite{weekes89}; \cite{reynolds93};
\cite{goret93}), the cut-off necessarily resides in the $\sim$100~GeV
energy range.  This is our primary motivation for this deep search for
pulsations from the Crab in the 100~GeV range.

The outer gap model by Romani 1996 also includes TeV emission via the
synchrotron-self-Compton mechanism which produces a peak spectral energy
density above 1~TeV. Such a mechanism could in principle explain the
detection of pulsed emission by the Durham group, which operates at an
energy threshold of 1~TeV, and still be consistent with the upper
limits reported at lower energies. For this reason we have applied
spectral analysis techniques to search for a gamma-ray Crab pulsar
signal over the energy band 250~GeV to 4~TeV.

\section{Observation and Analysis Techniques}
\label{section:tech}
The VHE observations reported in this paper utilize the atmospheric
Cherenkov technique (\cite{cawley95}) and the 10~m optical reflector
located at the Whipple Observatory on Mt. Hopkins in southern Arizona
(elevation 2.3~km) (\cite{cawley90}).  A camera, consisting of
photomultiplier tubes (PMTs) mounted in the focal plane of the
reflector, detects the Cherenkov radiation produced by gamma-ray and
cosmic-ray air showers from which an image of the Cherenkov light can
be reconstructed. For most of the observations reported here, the
camera consisted of 109~PMTs (each viewing a circular field of
0\fdg259 radius) with a total field of view of $3\degr$ in
diameter. In 1996 December, 42 additional PMTs were added to the
camera, increasing the field of view to 3\fdg3.

We characterize each Cherenkov image using a moment analysis
(\cite{reynolds93}). The roughly elliptical shape of the image is
described by the {\em length} and {\em width} parameters and its
location and orientation within the field of view are given by the
{\em distance} and $\alpha$ parameters, respectively. We also
determine the two highest signals recorded by the PMTs ({\em max1,
max2}) and the amount of light in the image ({\em size}). These
parameters are defined in Table~\ref{table:hillas} and are depicted in
Figure~\ref{figure:hillas}.  Gamma-ray events give rise to more
compact shower images than background hadronic showers and are
preferentially oriented towards the putative source position in the
image plane. By making use of these differences, a gamma-ray signal
can be extracted from the large background of hadronic showers.

\subsection{Selection Methods}  
\label{section:selection}
The standard gamma-ray selection method utilized by the Whipple
Collaboration is the Supercuts criteria (see
Table~\ref{table:supercuts}; cf., \cite{reynolds93};
\cite{catanese96}). These criteria were optimized on contemporaneous
Crab Nebula data to give the best sensitivity to point sources.  In an
effort to remove the background of events triggered by single muons
and night sky fluctuations, Supercuts incorporates pre-selection cuts
on the {\em size} and on {\em max1} and {\em max2}. While the
introduction of a pre-selection is desirable from the point of view of
optimizing overall sensitivity, it automatically rejects many showers
below $\sim400$~GeV. In the context of a search for pulsed emission
from the Crab pulsar, which must have a low energy cut-off to
accommodate existing upper limits, this is clearly
undesirable. Accordingly, a modified set of cuts
(Table~\ref{table:smallcuts}; cf., \cite{moriarty97}), developed to
provide optimal sensitivity in the $\sim200$~GeV to $\sim400$~GeV
region and referred to hereafter as Smallcuts, was used for the events
which failed the Supercuts pre-selection criteria. The most notable
difference between Smallcuts and Supercuts is the introduction of a
cut on the {\em length/size} of an image. Such a cut is effective at
discriminating partial arcs of Cherenkov light rings arising from
single muons, which become the predominant background at lower
energies. These images tend to be long compared to their intensity and
so may be rejected on the basis of the {\em length/size} ratio. When a
combination of Supercuts and Smallcuts is used, Monte Carlo
simulations indicate that this analysis results in an energy threshold
of $\sim250$~GeV. This threshold is the energy at which the
differential rate from a source with a spectral index equal to that of
the steady Crab Nebula reaches its peak. The collection area as a
function of gamma-ray energy is depicted in
Figure~\ref{figure:collection} and results in an effective collection
area of $2.7\times10^8~{\rm cm}^2$.  Details of the methods used to
estimate the energy threshold and effective area are given elsewhere
(\cite{mohanty98}).

The data from 1997 were analyzed with slightly modified cuts (see
Tables~\ref{table:supercuts},\ref{table:smallcuts}) which were
re-optimized after an upgrade to the Whipple camera which increased
the field of view. The greatest effect of the larger field of view was
that images appeared longer and at a greater distance from the center
of the field of view due to less image truncation than caused by the
smaller camera.

Supercuts was optimized to give the best point source sensitivity but
in doing so it rejects many of the larger gamma-ray events. Another
selection process, known as Extended Supercuts
(Table~\ref{table:extendedsupercuts}; cf., \cite{mohanty98}), was
utilized to facilitate a search for pulsed emission over the energy
band 250~GeV to 4~TeV. This method is quite similar to Supercuts but
scales the various cuts with the shower {\em size} and retains
approximately 95\% of gamma-ray events compared to approximately 50\%
of gamma-ray events passed by the Supercuts criteria. By applying a
lower bound on the {\em size} of an image, the energy threshold of the
analysis increases.  Figure~\ref{figure:collection} depicts the
collection area as a function of gamma-ray energy as derived by Monte
Carlo simulations for a lower bound on the {\em size} of an image of
500, 1000, 2000 and 5000 digital counts. These cuts impose energy
thresholds of 0.6, 1.0, 2.0 and 4.0~TeV respectively.

\subsection{Periodic Analysis}
The arrival times of the Cherenkov events were registered by a GPS
clock with an absolute resolution of 250~$\mu$s. An oscillator,
calibrated by GPS second marks (relative resolution of 100~ns), was
used to interpolate to a resolution of 0.1~$\mu$s. After an oscillator
calibration was applied, all arrival times were transformed to the
solar system barycenter by utilizing the JPL DE200 ephemeris as
described by Standish (1982). As the acceleration of the pulsar
relative to the solar system barycenter is negligible, the only additional
correction factor is due to the gravitational redshift. The conversion
of the coordinated universal time (UTC) as measured at the telescope,
to the solar system barycenter arrival time (TDB), is given by
\begin{equation}
t_{TDB} = t_{UTC} + \Delta_{TAI-UTC} + \Delta_{TDT-TAI} + 
          \Delta_{TDB-TDT} + \Delta_{REL}.
\end{equation}
The international atomic time (TAI) differs from UTC time by an
integral number of leap seconds. The terrestrial dynamical time (TDT)
is used as a timescale of ephemerides for observations from the
Earth's surface and differs from TAI by 32.184 s. The correction to
the Earth's surface requires the telescope's geocentric coordinates
and a model of the Earth's motion. The final correction applied,
$\Delta_{REL}$, accounts for the variation of the gravitational
potential around the Earth's orbit.

The corrected times were folded to produce the phases, $\phi_{j}$, of
the events modulo the pulse period according to
\begin{equation}
\phi_j = \phi_0 + \nu(t_j - t_0) + \frac{1}{2}\dot{\nu} (t_j - t_0)^2,
\end{equation}
where $\nu, \dot{\nu}$ are the frequency and first frequency
derivative at the epoch of observation $t_0$.  For each source run the
valid frequency parameters were derived from the J2000 ephemeris
obtained from Jodrell Bank where the Crab pulsar is monitored on a
monthly basis.

To check the Whipple Observatory timing systems an \underline{{\em
optical}} observation of the Crab pulsar was undertaken on the nights
of 1996 December 2 (UT), 1996 December 18 (UT) and 1997 March 11 (UT),
using the 10~m reflector with a photometer at its focus
(\cite{srinivasan97}). The signal from the photometer was recorded by
the data acquisition electronics and timing system of the telescope
thereby providing a direct test of the instrument's timing
characteristics. The phase analysis of the event arrival times,
depicted in Figure~\ref{figure:optphase}, yielded a clear detection of
the optical signal from the Crab pulsar in phase with the radio
pulse. This demonstrates the validity of the timing, data acquisition
and barycentering software in the presence of a pulsed signal.

\section{Observations and Results} 
\label{section:results}
The position of the Crab pulsar was observed between 1995 January and
1997 March. The traditional mode of observing potential periodic
sources with the Whipple Observatory gamma-ray telescope is to track
the putative source location continuously for runs of 28 minute
duration. After filtering runs for bad weather and instrumental
problems, the data set consists of 159 runs for a total source
observing time of 73.4 hrs.  The radio position (J2000) of the Crab
pulsar ($\alpha$ = 05$^h$ 34$^m$ 31.949$^s$, $\delta$= 22$\degr$
00$^{\prime}$ 52.057$^{\prime\prime}$) was assumed for the subsequent
timing analysis.

The numbers of events passing the selection criteria described above
are given in Table~\ref{table:selectedevents}. The phases of these
events, shown in Figure~\ref{figure:tevphase}, are used for periodic
analysis. We find no evidence of pulsed emission at the radio period.
To calculate upper limits for pulsed emission we have used the pulse
profile seen at lower energies by EGRET. That is, we assume emission
occurs within the phase ranges of both the main pulse, phase
0.94-0.04, and the intrapulse, phase 0.32-0.43 (\cite{fierro98}). The
number of events with phases within these intervals constitutes the
number of candidate pulsed events, $N_{on}$. $N_{off}$, an estimate of
the numbers of background events, is obtained by multiplying the
number of events with phases outside these pulse intervals by the
ratio of ranges spanned by the pulse and non-pulse regions. The
results are given in Table~\ref{table:selectedevents}. The statistical
significance of the excess is calculated using the maximum likelihood
method of Li~\&~Ma (1983). The 99.9\% confidence level upper limits
calculated using the method of Helene (1983) are given in
Table~\ref{table:upperlimits}.

Several reports of pulsed emission from the Crab pulsar at very high
energies claim to have seen evidence of episodic emission on time
scales of several minutes. For this reason we have performed a
run-by-run search for periodic emission from the Crab pulsar based
on the above pulse profile. The statistical significance of excess
events for each observation and the corresponding distribution of
significance for the lowest and middle energy ranges are given in
Figure~\ref{figure:episodic}. In each energy band the distribution of
significance is consist with the statistical expectation for zero
excess.

\section{Discussion}
\label{section:discuss}
Data taken with the Whipple Observatory's 10~m gamma-ray telescope
have been used to search for pulsations from the Crab pulsar above
250~GeV. We find no evidence of pulsed emission at the radio period
and upper limits on the integral flux have been given.

To model the pulsed gamma-ray spectrum, a function of the form
\begin{equation}
dN/dE = KE^{-\gamma}e^{-E/E_o}
\label{equation:powercutoff}
\end{equation}
was used, where $E$ is the photon energy, $\gamma$ is the photon
spectral index and $E_o$ is the cut-off energy. The source spectrum in
the EGRET energy range is well fitted by a power law with a photon
spectral index of $-2.15\pm0.04$ (\cite{nolan93}). The pulsed upper
limit above 250~GeV reported here is $\sim3$ orders of magnitude below
the flux predicted by the EGRET power law.
Equation~\ref{equation:powercutoff} was used to extrapolate the EGRET
spectrum to higher energies constrained by the TeV upper limit
reported here and indicates a cut-off energy $E_o \leq 60$ GeV for
pulsed emission (see Figure~\ref{figure:crabsd}).

As indicated in \S~\ref{section:selection}, the energy threshold
of the technique is derived assuming a source with a spectral index
equal to that of the steady Crab Nebula. With the above model, this
assumption is invalid. If we assume a source spectrum as given by
Equation~\ref{equation:powercutoff} and define energy threshold and
effective collection area as stated in \S~\ref{section:selection}
we simultaneously solve for an energy threshold of 180 GeV and energy
cut-off of 60 GeV. The derived cut-off energy is the same as that obtained
assuming a Crab Nebula spectrum, and indicates the robustness
of defining the energy threshold of the technique in this way.

The sharpness of the spectral cut-off of the emission models depicted
in Figure~\ref{figure:crabsd} provides a good discriminant. The status
of current observations and the derived cut-off given above indicates
that the cut-off must lie in the 10-60~GeV range. However, the upper
limits reported here are well above the flux predicted by the polar
cap and outer gap models and offer no discrimination between them. In
contrast, the outer gap model of Romani (1996) predicts TeV emission
via the synchrotron-self-Compton mechanism. The flux produced via this
mechanism is dependent on the density and spectrum of primary electrons
and positrons in the gap, as well as the density of local soft photon
fields. The predicted pulsed TeV flux for a young gamma-ray pulsar is
somewhat less than 1\% of the pulsed GeV flux. The results reported here
derive an upper limit to this fraction of $<0.07$\%.

\acknowledgments We acknowledge the technical assistance of K. Harris,
T. Lappin, and E. Roache.  We thank A. Lyne and R. Pritchard for
providing the radio ephemeris of the Crab pulsar. This research is
supported by grants from the U.S.  Department of Energy, NASA, the
Irish National Research Support Fund Board and by PPARC in the United
Kingdom.

\newpage
\figcaption[fig1.eps]{Images of Cherenkov light produced by
gamma-ray and cosmic-ray induced air showers are parameterized using a
moment analysis to describe the image shape and orientation in the
image plane.\label{figure:hillas}}

\figcaption[fig2.eps]{The collection areas for
Supercuts combined with Smallcuts (SC+SM) and Extended Supercuts (ESC)
combined with a lower bound on image {\em size}, which selects only
the higher energy events.\label{figure:collection}}

\figcaption[fig3.eps]{Optical observations of the Crab pulsar with
the Whipple 10 m telescope. The datasets show a clear detection of the
Crab optical pulsations. Phase 0.0 (in each case) corresponds to the
extrapolated arrival of the radio peak closest to the epoch of
observations derived from the Jodrell Bank timing solution. The counts
in each case were normalized to the average of the respective
observation. a) Dataset taken on 1996 December 2 (UT); b) Dataset
taken on 1996 December 18 (UT); c) Dataset taken on 1997 March 11
(UT); d) Addition of the above datasets (in phase). The dashed lines
depicts the EGRET main and intrapulse phase ranges.\label{figure:optphase}}

\figcaption[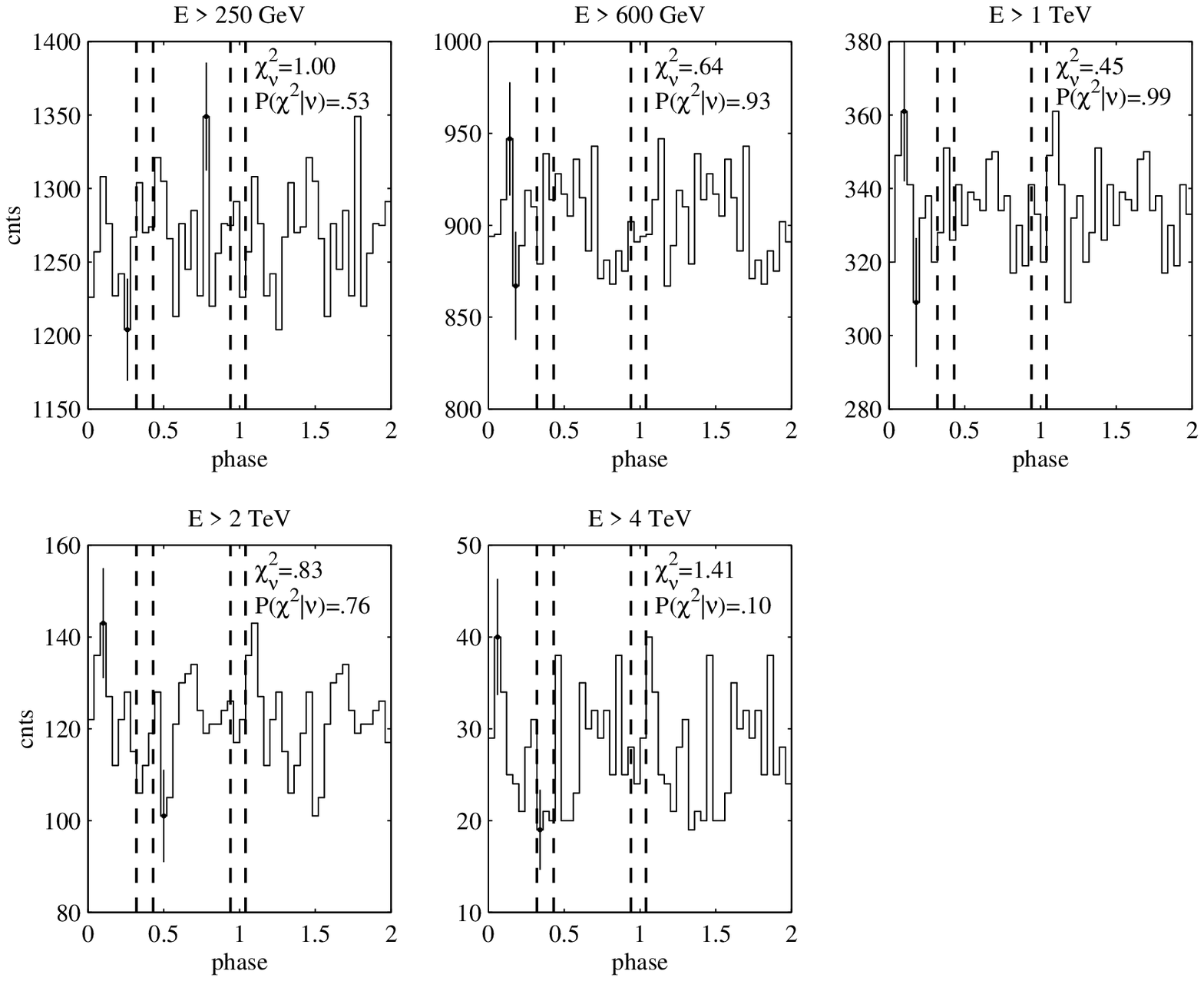]{Search for TeV gamma rays from the Crab
pulsar. The dashed lines depict the EGRET main pulse and intrapulse
phase ranges. Error bars have been included on bins with the maximum
and minimum number of counts. The $\chi^2$ probability that each
distribution is consistent with it's mean is given in each panel. We
find no evidence of pulsed emission at the radio
period.\label{figure:tevphase}}

\figcaption[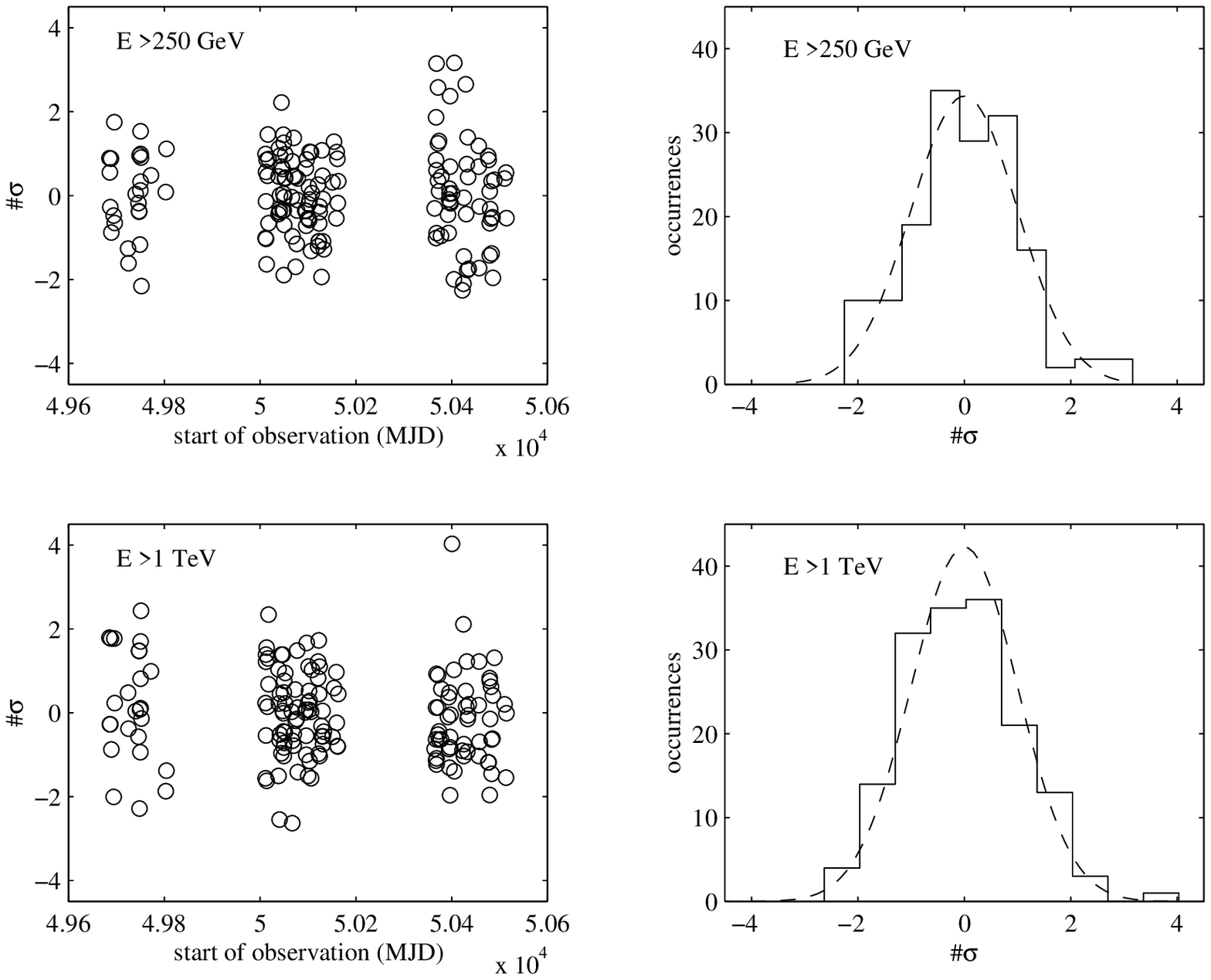]{The panels on the left depict the
statistical significance of excess events for each observation for
energy thresholds of 250 GeV and 1 TeV. The right panels depict the
corresponding distributions of significances as solid lines. The
dashed curves shows the statistical expectation for zero excess.
\label{figure:episodic}}

\figcaption[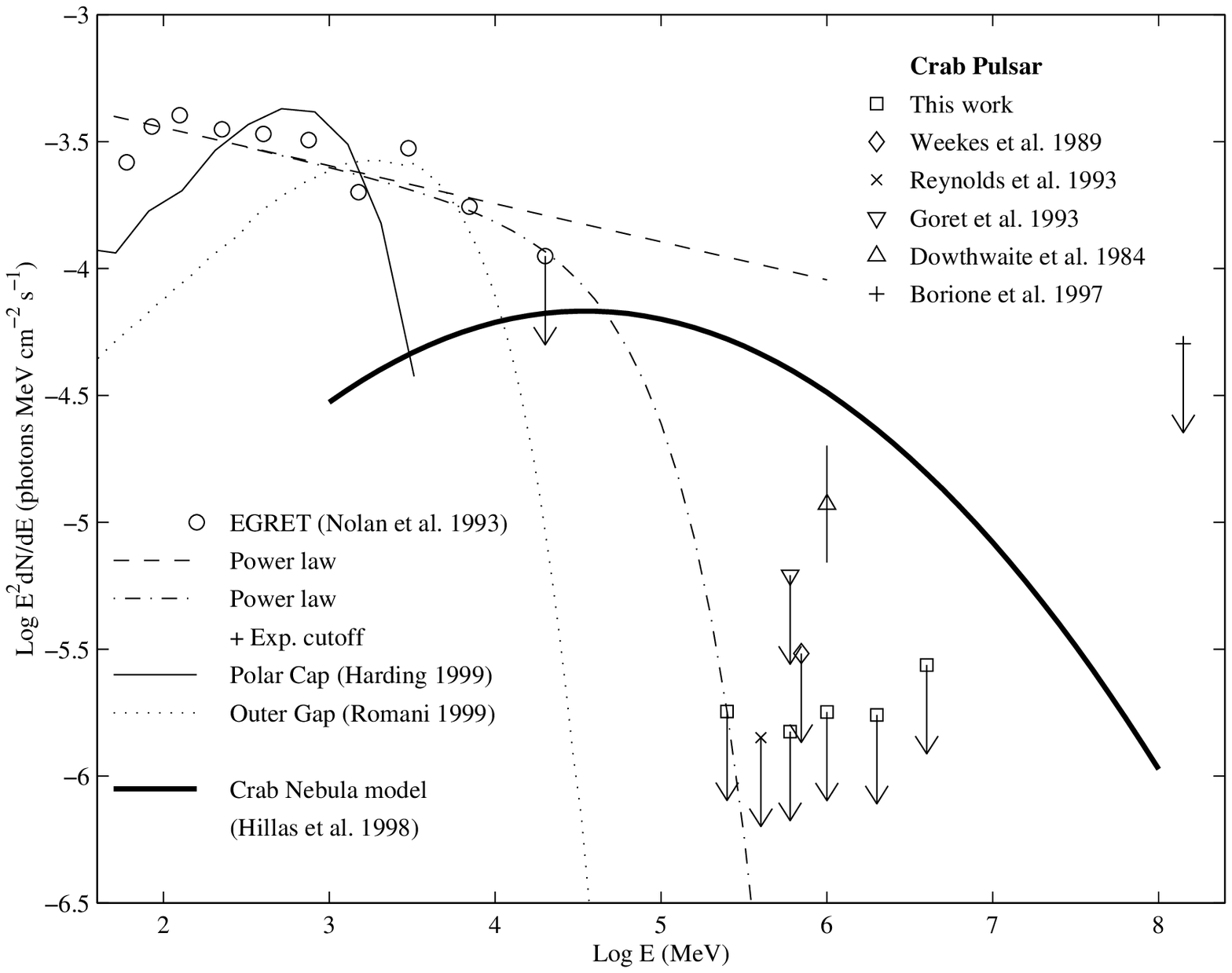]{The pulsed photon spectrum of the Crab
pulsar. The EGRET data points are from Nolan et al. 1993. The thin
solid line is the polar cap model fit to the EGRET data
(\cite{harding99}). The dotted line is the outer gap model for the
Vela pulsar (scaled to match the EGRET Crab pulsar flux at peak
intensity) and is included to indicate the shape of the cut-off this
model predicts (\cite{romani99}). The dashed line represents the
power-law fit to the EGRET data (\cite{nolan93}). The dot-dashed line
represents Equation~\ref{equation:powercutoff} with a cut-off energy
$E_o = 60$ GeV. The upper limits for pulsed emission presented in this
paper are represented by the open squares. The thick solid curve
depicts the model of unpulsed GeV - TeV emission from the Crab Nebula
(\cite{hillas98}).\label{figure:crabsd}}

\newpage

\begin{deluxetable}{ll}
\tablecaption{Definition of the image parameters, which are used to
characterize the image shape and orientation (see
Figure~\protect\ref{figure:hillas}).\label{table:hillas}}
\tablehead{\colhead{parameter} & \colhead{definition}}
\startdata
{\em max1}:     & largest signal recorded by the PMTs \\
{\em max2}:     & second largest signal recorded by the PMTs \\
{\em size}:     & sum of all signals recorded \\
{\em width}:    & the root mean square (RMS) spread of light along the \\
                & minor axis of the image; a measure of the lateral \\
                & development of the shower.      \\
{\em length}:   & the RMS spread of light along the major axis of the image;\\
                & a measure of the vertical development of the shower.     \\
{\em distance}: & the distance from the centroid of the image to the center \\ 
                & of the field of view. \\
$\alpha$:       & the angle between the major axis of the image and a line \\
                & joining the centroid of the image to the center of the\\
                & field of view. \\
\enddata
\end{deluxetable}

\begin{deluxetable}{cc}
\tablecaption{Supercuts gamma-ray selection criteria.\label{table:supercuts}}
\tablehead{\colhead{Supercuts (1995/1996)} & 
           \colhead{Supercuts (1997)}}
\startdata
\cutinhead{pre-selection criteria}                      \\
{\em max1} $>$ 100 d.c.\tablenotemark{a}&	{\em max1} $>$ 95 d.c.	\\
{\em max2} $>$ 80 d.c.		        &	{\em max2} $>$ 45 d.c.	\\
{\em size} $>$ 400 d.c. 	        &	{\em size} $>$ 0 d.c.	\\
\cutinhead{gamma-ray selection}                         \\
0\fdg073 $<$ {\em width}    $<$ 0\fdg15 &
0\fdg073 $<$ {\em width}    $<$ 0\fdg16 \\
0\fdg16  $<$ {\em length}   $<$ 0\fdg30 &
0\fdg16  $<$ {\em length}   $<$ 0\fdg33 \\
0\fdg51  $<$ {\em distance} $<$ 1\fdg10 &
0\fdg51  $<$ {\em distance} $<$ 1\fdg17 \\
$\alpha < 15\degr$		&	$\alpha < 15\degr$ \\
\tablenotetext{a}{d.c. = digital counts (1.0 d.c. $\approx$ 1.0 photoelectron).}
\enddata
\end{deluxetable}

\begin{deluxetable}{cc}
\tablecaption{Smallcuts gamma-ray selection criteria applied to events which
failed the Supercuts pre-selection criteria.\label{table:smallcuts}}
\tablehead{\colhead{Smallcuts (1995/1996)} & 
           \colhead{Smallcuts (1997)}}
\startdata
{\em max1} $>$ 40 d.c.		& {\em max1} $>$ 40 d.c. 		\\
{\em max2} $>$ 40 d.c.		& {\em max2} $>$ 40 d.c. 		\\
{\em size} $>$  0 d.c. 		& {\em size} $>$  0 d.c.		\\
{\em length/size} $< 8.3\times10^{-4}$ $\degr$/d.c. &
{\em length/size} $< 8.3\times10^{-4}$ $\degr$/d.c. \\
0\fdg073 $<$ {\em width}    $<$ 0\fdg13   & 
0\fdg073 $<$ {\em width}    $<$ 0\fdg13   \\
0\fdg16  $<$ {\em length}   $<$ 0\fdg30   &
0\fdg16  $<$ {\em length}   $<$ 0\fdg33   \\
0\fdg51  $<$ {\em distance} $<$ 1\fdg10   &
0\fdg51  $<$ {\em distance} $<$ 1\fdg17   \\
$\alpha < 15\degr$              & $\alpha < 15\degr$
\enddata
\end{deluxetable}

\begin{deluxetable}{c}
\tablecaption{Extended Supercuts.\label{table:extendedsupercuts}}
\tablehead{Extended Supercuts (1995/1997)}
\startdata
{\em max1} $>$ 70 d.c.	\\
{\em max2} $>$ 70 d.c.	\\
{\em size} $>$ 500,1000,2000,5000 d.c. 	\\
$\left| width + 0\fdg022 - 0\fdg023 \ln(size)\right| < 0\fdg048$    \\
$\left| length - 0\fdg114 - 0\fdg020 \ln(size)\right| < 0\fdg068$   \\
0\fdg6    $<$ {\em distance} $<$ 1\fdg0 \\
$\alpha - 9\fdg16 + 0\fdg558 \ln(size) < 13\fdg5$
\enddata
\end{deluxetable}

\begin{deluxetable}{lccc}
\tablecaption{Selected events for periodic analysis. $N_{on}$ are the number
of events with phases within the EGRET pulse profile and $N_{off}$ are
the background estimated from events falling outside the EGRET pulse
profile.\label{table:selectedevents}}
\tablehead{\colhead{Selection} & 
           \colhead{$N_{on}$} & 
           \colhead{$N_{off}$} & 
           \colhead{Significance ($\sigma$)}}
\startdata
Supercuts + Smallcuts                   & 6696     & 6636     & 0.65\\
Extended Supercuts ({\em size} $>$ 500) & 4709     & 4748     & -0.50\\
Extended Supercuts ({\em size} $>$ 1000)& 1738     & 1762     & -0.51\\
Extended Supercuts ({\em size} $>$ 2000)& 602      & 649      & -1.67\\
Extended Supercuts ({\em size} $>$ 5000)& 125      & 150      & -1.88\\
\enddata
\end{deluxetable}

\begin{deluxetable}{lcc}
\tablecaption{Integral Flux Upper limits.\label{table:upperlimits}}
\tablehead{\colhead{Selection Method} &
           \colhead{Periodic Emission (cm$^{-2}$s$^{-1}$)$\times$10$^{-13}$} & 
           \colhead{Threshold (TeV)}}
\startdata
Supercuts + Smallcuts             & $<$48.2             & $\geq$ 0.25 \\
Extended Supercuts (size $>$ 500) & $<$16.7             & $\geq$ 0.6  \\
Extended Supercuts (size $>$ 1000)& $<$12.0             & $\geq$ 1.0  \\
Extended Supercuts (size $>$ 2000)& $<$5.9              & $\geq$ 2.0  \\
Extended Supercuts (size $>$ 5000)& $<$4.6              & $\geq$ 4.0  \\
\enddata
\end{deluxetable}


\clearpage

\plotone{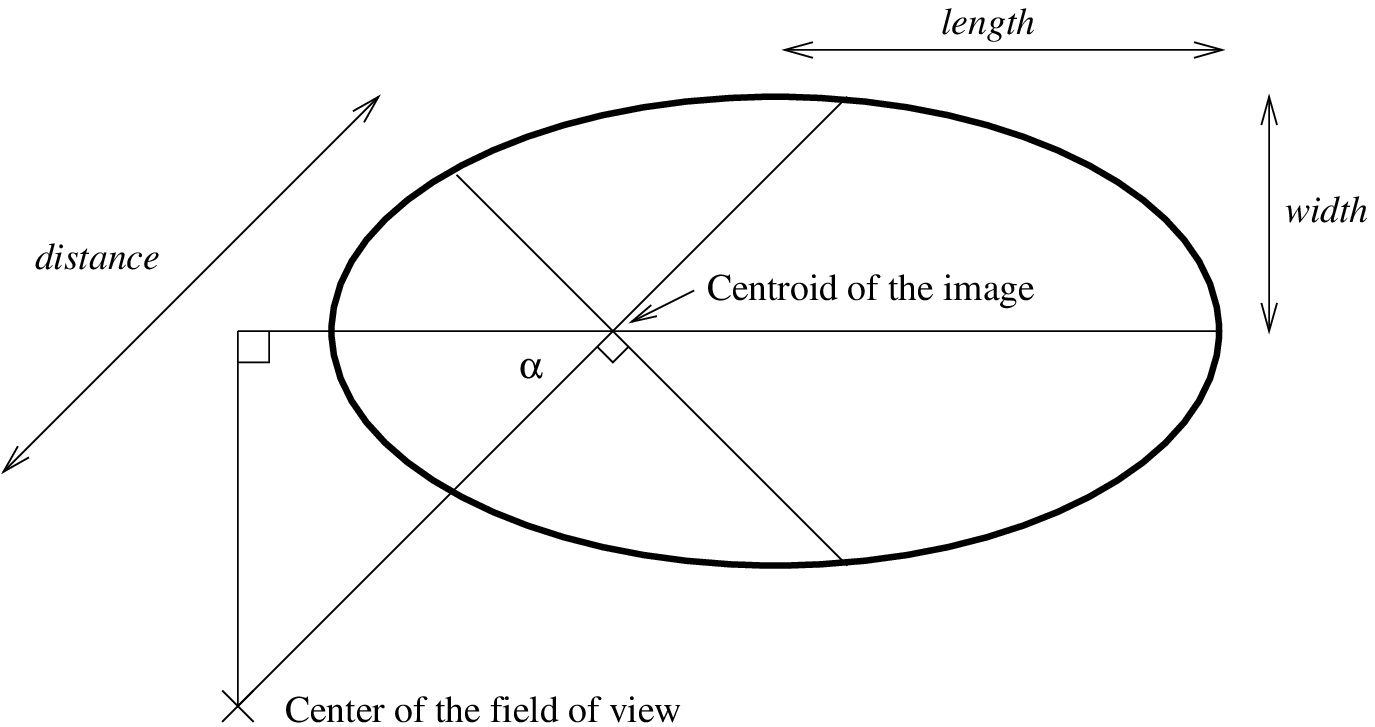}

\clearpage

\plotone{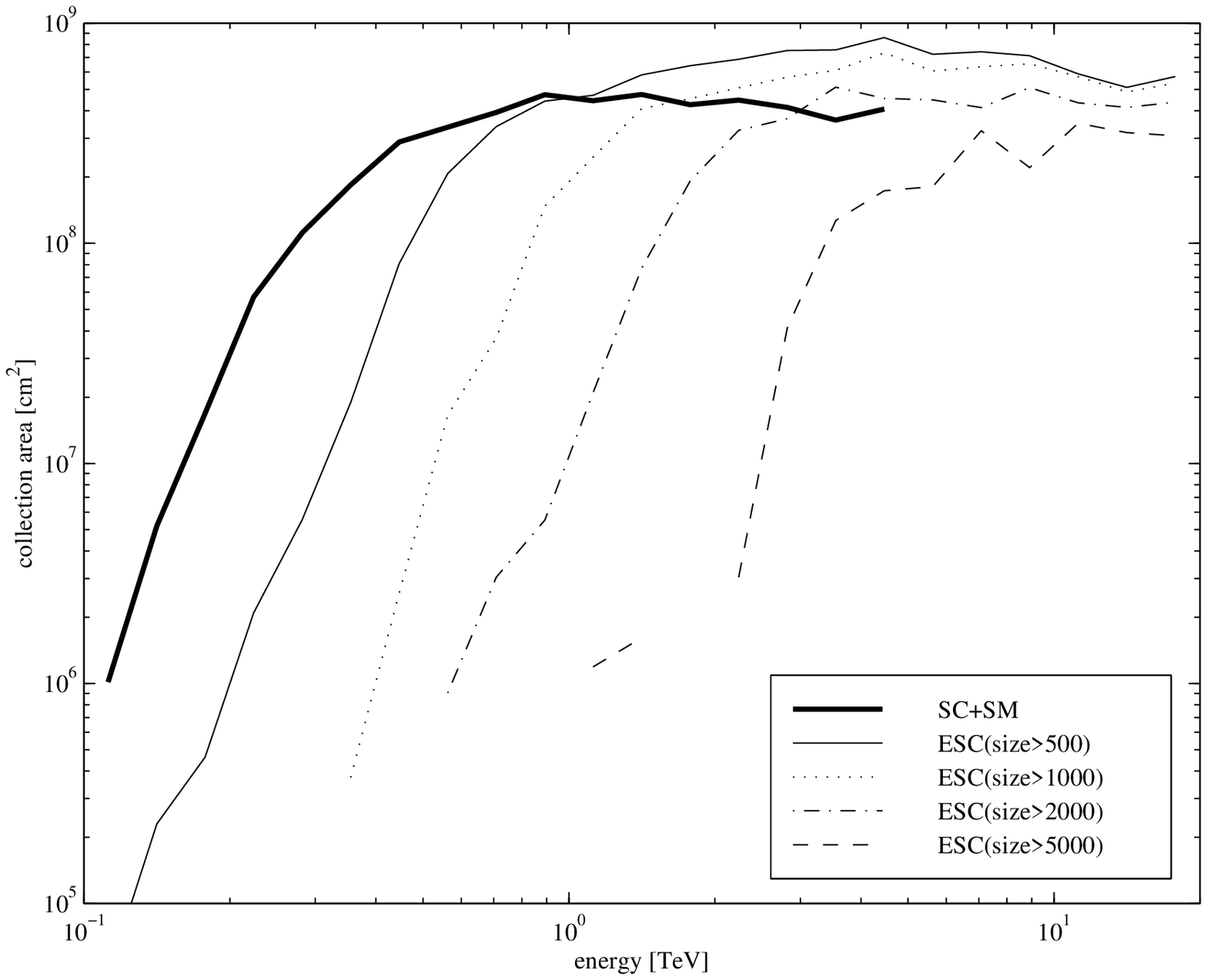}

\clearpage

\plotone{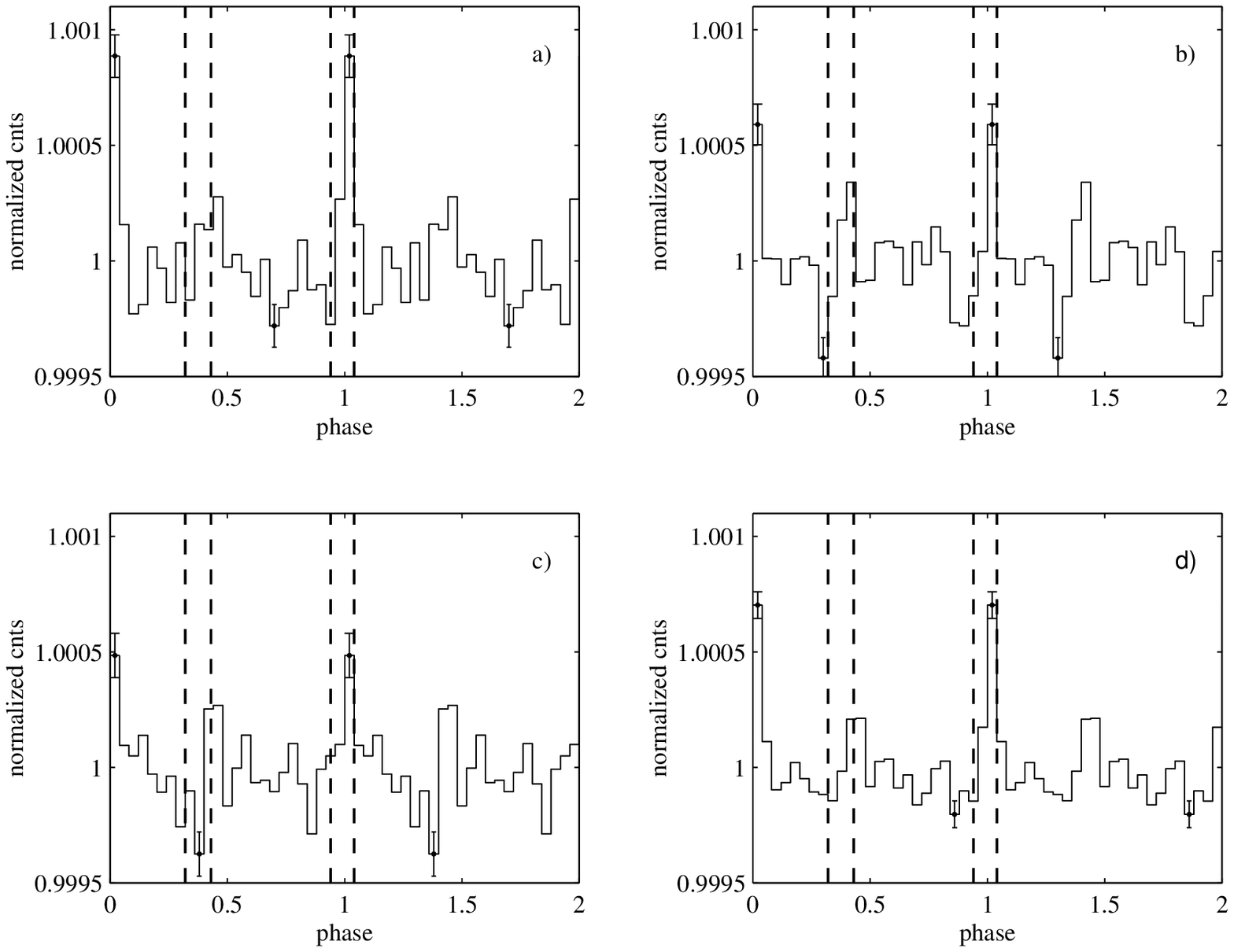}

\clearpage

\plotone{f4.eps}

\clearpage

\plotone{f5.eps}

\clearpage

\plotone{f6.eps}
\end{document}